\documentclass{PoS}
\usepackage{amsmath}
\usepackage{amsfonts}
\usepackage{amssymb}
\usepackage{array}
\usepackage{feynmp}
\usepackage[pdftex]{graphicx}
\usepackage{cite}

\title{Rho decay width from the lattice}

\ShortTitle{Rho decay width from the lattice}

\author{\speaker{J. Frison}$^a$, S. Durr$^{b,c}$, Z. Fodor$^{b,c,d}$, C. Hoelbling$^b$, S.D. Katz$^{b,d}$, S. Krieg$^{b,c}$, T. Kurth$^b$, L. Lellouch$^a$, T. Lippert$^{b,c}$, A. Portelli$^a$, A. Ramos$^a$ and K.K. Szabo$^b$ (Budapest-Marseille-Wuppertal~Collaboration)
\\
\llap{$^a$}Centre de Physique Th\'eorique
\thanks{CPT is research unit UMR 6207 of the CNRS and of the universities Aix-Marseille II, Aix-Marseille I and
Sud Toulon-Var, and is affiliated with the FRUMAM.}
, Case 907, CNRS Luminy, F-13288 Marseille Cedex 9, France.\\
\llap{$^b$}Bergische Universit\"at Wuppertal, Gaussstr. 20, D-42119 Wuppertal, Germany.\\
\llap{$^c$}J\"ulich Supercomputing Centre, Forschungszentrum J\"ulich, D-52425 J\"ulich, Germany.\\
\llap{$^d$}Institute for Theoretical Physics, E\"otv\"os University, H-1117 Budapest, Hungary.\\
       E-mail: \email{frison@cpt.univ-mrs.fr}
}

\abstract{
While the masses of light hadrons have been extensively studied in lattice QCD simulations, there exist only a few exploratory calculations of the strong decay widths of hadronic resonances. We will present preliminary results of a computation of the rho meson width obtained using $N_f=2+1$ flavor simulations. The work is based on L\"uscher's formalism and its extension to moving frames.}

\FullConference{The XXVIII International Symposium on Lattice Field Theory\\
                June 14-19,2010\\
                Villasimius, Sardinia Italy}

\begin{document}

The light hadron masses have been extensively and quite successfully studied in lattice QCD simulations\cite{BMW08,christian10}. However, the study of strong decays remains a challenge that only a few exploratory
calculations have addressed so far\cite{aoki07,xufeng}.

Until recently this issue could be ignored : firstly because the sea quarks required for a multi-body decay are not present in the quenched approximation, and then because energy conservation leaves little or no phase space in those decays for unphysically large light quark masses. But as we get closer to the physical point and improve our precision, this issue has to be addressed for us to be able to reliably determine masses of resonant states.
Additionally, it is exciting way to test the characteristic dynamical effects of sea quarks.

A finite-volume formalism has been developped by L\"uscher\cite{luscher86-1,luscher86-2,luscher90,luscher91-1,luscher91-2}, and extended to moving frames\cite{gottlieb96}, which describes the modification of the quantization condition of scattering states momenta under interactions. We use this framework to compute the rho decay width on a subset of the Budapest-Marseille-Wuppertal collaboration configurations\cite{physpoint-A,physpoint-B}, which features improved Wilson fermions with $N_f = 2 + 1$ flavors of sea quarks.

\section{Introducing the two-pions levels}
\subsection{Basics of avoided crossings}
\begin{figure}[t]
\centering
\includegraphics[width=0.4\linewidth]{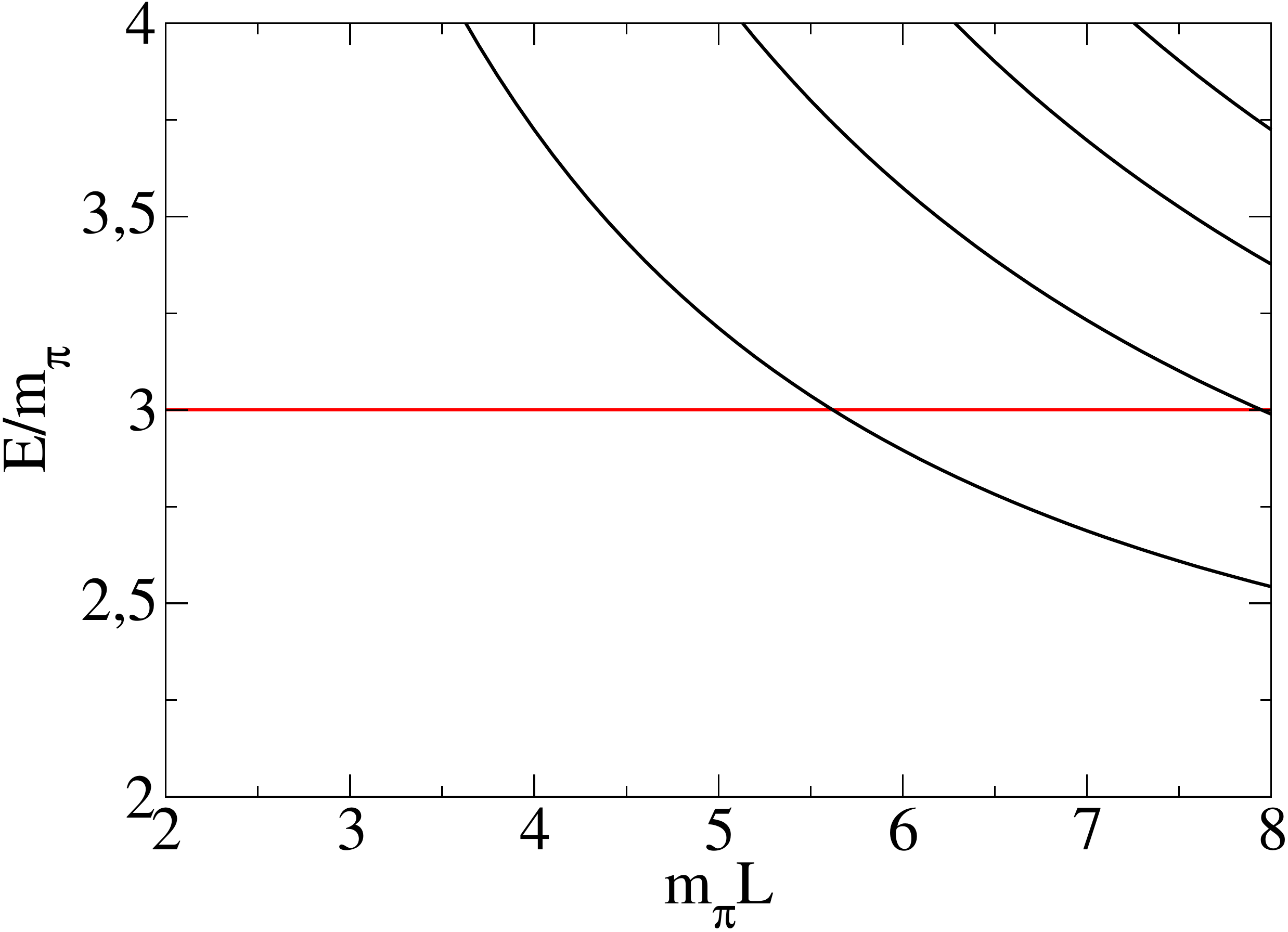}\hspace{1cm}
\includegraphics[width=0.4\linewidth]{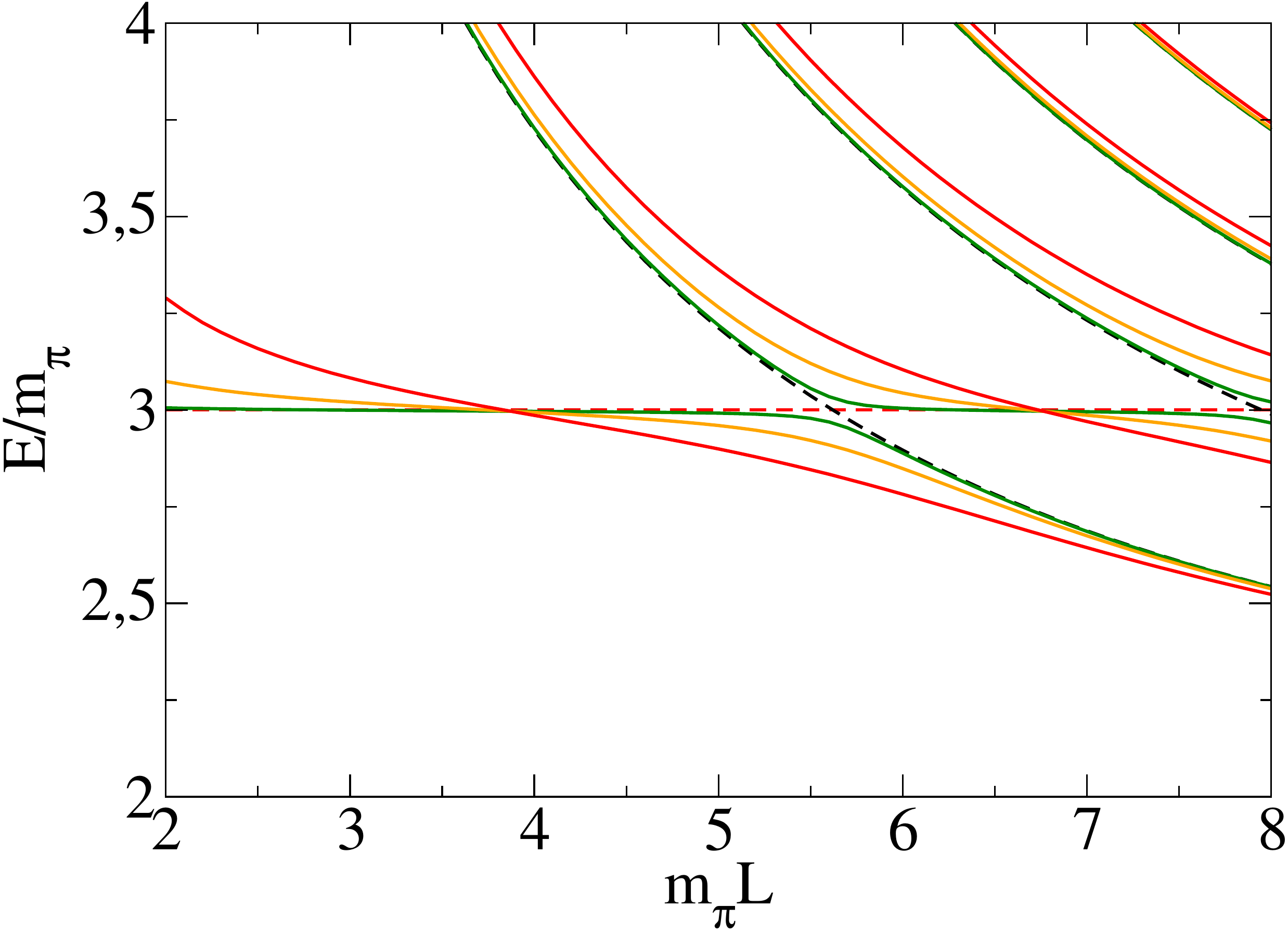}
\caption{The left side shows the spectrum in the free case as a function of $m_\pi L$, for $m_\rho/m_\pi=3$ and ignoring exponential corrections. This free spectrum is reproduced in dash lines on the right side and then we show its deformation with different couplings. The model is the one used in section 1.3, with the coupling getting stronger from green to red.} 
\label{fig:thspect}
\end{figure}

If $\rho$ were not coupled to $\pi\pi$, the spectrum as a function of box size $L$ would only consist of :
\begin{itemize}
\item a $\rho$ state, whose energy would be a constant $m_\rho$ up to exponentially suppressed finite-volume corrections\cite{luscher86-1}, and
\item free two-pions states, whose momenta $\vec{k}=(2\pi/L)\vec n$, and energies $E=\sqrt{m_\pi^2+\vec k^2}$ are quantized.
\end{itemize}
For some particular box sizes the $\rho$ states crosses one of the free two-pions states. At these points we would have a degeneracy but nothing particular would happen. But if now we turn on interactions, the situation becomes very different: we know from quantum mechanics that two eigenstates cannot cross. The interaction mixes $\rho$ and $\pi\pi$ into new eingenstates which exhibit an avoided level crossing phenomenon (Fig.~\ref{fig:thspect}). These avoided crossings therefore contain information on the coupling between the states, and hence on the $\rho$ width.

\subsection{L\"uscher's Formula}
\begin{figure}[t]
\centering
\begin{tabular}{>{\raggedleft}m{8cm}>{\raggedright}m{4cm}}
\includegraphics[width=\linewidth]{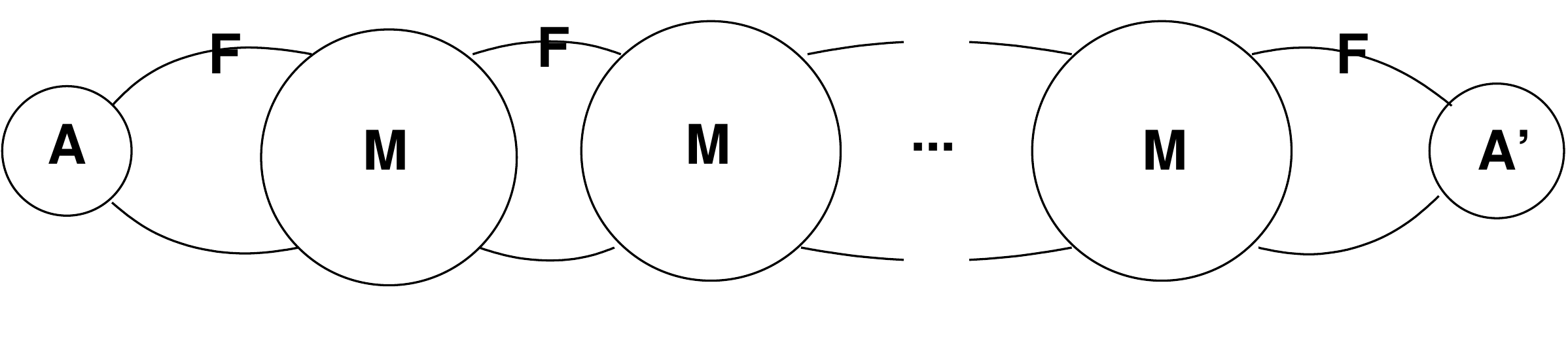} &   \LARGE{$= AF\frac{1}{1-MF}A'$}
\end{tabular}
\caption{
Here we represent the finite-volume correction of an arbitrary correlator, computed under the inelastic threshold up to exponentially small corrections. $M$ is the infinite-volume amputated four-point function, while $F$ is the difference between finite-volume loops and infinite-volume ones. $M$ is obtained as a resummation of a volume-independant series of $2$-particle-irreducible kernels.}
\label{fig:luscherBS}
\end{figure}
In a finite box we expect the pions to interact then propagate over small distance, then interact again, and so on. This can be represented as an expansion in a series of two-particles-irreducible kernels as shown in Fig.~\ref{fig:luscherBS}. When this expansion is resummed, the interactions move $F$ poles (free two-pions states) to $1-MF$ zeros\cite{kim05}. The zeros of $1-MF$ correspond to the solutions of L\"uscher's formula, which was first derived in \cite{luscher86-2}.

L\"uscher's formula is only valid below the $4\pi$ inelastic threshold. Expressed in a form exhibiting its role of quantization condition, it reads (up to exponentially small terms in $m_\pi L$) :
\begin{equation}
\Phi(q)=n\pi-\delta(q) ,
\end{equation}
using the reduced momentum $q=kL/2\pi$ and the $\delta$ phase shift in the $I=J=1$ channel. $\Phi$ is a known kinematical function of $q$, expressing the breaking of Lorentz invariance by the cubic box.

\subsection{The $g$-coupling model}
\label{sec:gmodel}
L\"uscher's equation gives us the phase shift for a few discrete values of the momentum, determined by the parameters of the lattice. It would be difficult and very costly numerically to significantly increase the number of such momenta. Therefore we need a model to reconstruct the phase shift from a few points. Moreover, this model will be useful to make the necessary extrapolations ($m_\pi\to m_\pi^{phys}$, $a\rightarrow 0$, $L\rightarrow\infty$, ... ). Following \cite{luscher91-2} we use both an effective range approximation and an effective lagrangian ${\cal L}_{eff}=g\epsilon_{abc}\rho_ \mu^a\pi^b\partial^\mu\pi^c$.

The effective range approximation parametrizes the phase shift in terms of a Taylor expansion around $k_\rho=\sqrt{E^2/4-m_\pi^2}$, where $k_\rho$ is the momentum of $\rho's$ decay products in infinite-volume:
\begin{equation}
\label{eq:effrange}
\frac{k^3}{W}\cot\delta=b(k^2-k_\rho^2) .
\end{equation}

The second ingredient allows us to parametrize the coupling of the $\rho$ to $\pi\pi$ states in terms of a constant coupling $g$ that should have a rather small dependance on the quark mass. On the other hand the width $\Gamma$ is strongly dependant on kinematics :\\
\begin{equation}
\label{eq:gammafromg}
\Gamma_\rho=\frac{g^2}{6\pi}\cdot\frac{k_\rho^3}{m_\rho^2} .
\end{equation}
Using experimental values for the masses and the width it yields $g\simeq 6.0$.\\

\section{Results with a single interpolating operator}
The formulae \ref{eq:effrange} and \ref{eq:gammafromg} can be combined into a very simple form, giving the correction to get the $\rho$ mass from the energy $E$ of an arbitrary eigenstate in the $I=J=1$ channel :
\begin{equation}
\label{eq:simplecorrections}
m_\rho^2=E^2-\frac{g^2}{6\pi}\frac{k^3}{E}\cot\Phi(kL/2\pi) \qquad\mathrm{where}\;k=\sqrt{\frac{E^2}{4}-m_\pi^2}
\end{equation}

In this expression, the unknowns are $m_\pi$ and $g$. Instead of using two energy levels at a given quark mass, one can extract the ground state energy level at different quark masses. But the kinematics required in that case are contradictory: on the one hand we would get no signal for the width if the state is far below the first free two-pion state; on the other hand, near the first crossing one cannot disentangle the contributions of the $\rho$ mass and width to the measured energy level.

In \cite{BMW08} we considered only situations on which the ground state obtained with a single operator is far from the first crossing. Thus, in combined fits of our 6-stout data, we obtained a precise determination of $m_\rho$ ($m_\rho\sim5\%$), but a much less precise determination of $g$ : $g=9.5\pm4.6$.

\section{Using two interpolating operators}
\subsection{Generalized Eigenvalue Problem}
Now we are going to quickly present the principle of extraction of several energies by the variationnal method. Let us first assume that we can compute observables only affected by $N$ eigenstates. We will show that energies can be computed from a set of $N\times N$ cross-correlators
\begin{equation}
C_{ij}(t)=\left<0\mid{\cal O}_i(t){\cal O}_j(0)\mid 0\right>,\quad\mathrm{with}\;i,j=1\dots N .
\end{equation}
We first decompose them on the energy eigenstates and express them in a compact matrix form :
\begin{eqnarray}
C_{ij}(t)&=&\sum_n \left<0\mid{\cal O}_i\mid n\right>e^{-E_nt}\left<n\mid {\cal O}_j\mid 0\right>\\
&=&\sum_{m,n} (V^\dagger)_{i,m} \cdot D_{m,n}(t) \cdot V_{n,j} ,
\end{eqnarray}
so $V_{n,j}$ is the $n$th-state content of the ${\cal O}_i$ operator and $D$ is a diagonal matrix containing the exponentials of the energies. Now we immediatly see that $C(t)C^{-1}(t_0)$ is diagonal in the eigenstates basis :
\begin{equation}
C(t)C^{-1}(t_0) = V^\dagger D(t) V \left( V^\dagger D(t_0) V \right)^{-1} = (V^\dagger)D(t-t_0)(V^\dagger)^{-1} ,
\end{equation}
and its eigenvectors give the energies through :
\begin{equation}
\lambda_i = e^{-E_i(t-t_0)} .
\end{equation}
This is exact if and only if the operators are linearly independent (so we can invert the $V$s). Note that in the case $N=1$ this is simply the effective mass method.

In practice we can only compute the matrix of operators obtained from a few $N$ operators. Thus only the $N$ low-lying levels are asymptotically known, and higher-levels are treated as a contamination at short times. The energies are computed with errors of order $\exp[(E_{N}-E_{N+1})t]$ and $\exp[(E_{N}-E_{N+1})t_0]$, so $t$ {\it and} $t_0$ must be large ($t-t_0$ is not very important, since it only enters in polynomial prefactors). A more precise study of high-levels corrections on some derived quantities is made in \cite{luscher90}.

\subsection{Operators and contractions}
To implement the generalized eigenvalue approch described above, we consider two operators. The first is the point $\rho$ meson operator
\begin{equation}
\rho_i = \bar u \gamma_i u - \bar d \gamma_i d .
\end{equation}
The second operator must be non-local to have a sufficiently independent coupling to scattering states. We construct it from local pion operators :
\begin{equation}
{\cal \pi\pi}_i(\vec p,\vec q) = (p_i-q_i)\left[\pi^+(\vec p)\pi^-(\vec q)-\pi^-(\vec p)\pi^+(\vec q)\right] .
\end{equation}
Then, the cross-correlator is computed by contracting ``stochastic propagators''. Following \cite{aoki07} we use two kinds of ``stochastic propagators'' :
\begin{eqnarray}
Q(\vec x,t\mid\vec q,t_s,\xi_j)&=&\sum_{\vec y}D^{-1}(\vec x,t;\vec y,t_s)\cdot [e^{i\vec p\cdot\vec y}\xi_j(\vec y)]\\
W(\vec x,t\mid\vec k,t_1\mid\vec q,t_s)&=&\sum_{\vec z}D^{-1}(\vec x,t;\vec z,t_1)\cdot[e^{i\vec k\cdot\vec z}\gamma_5Q(\vec z,t_1\mid \vec q,t_s)]
\end{eqnarray}
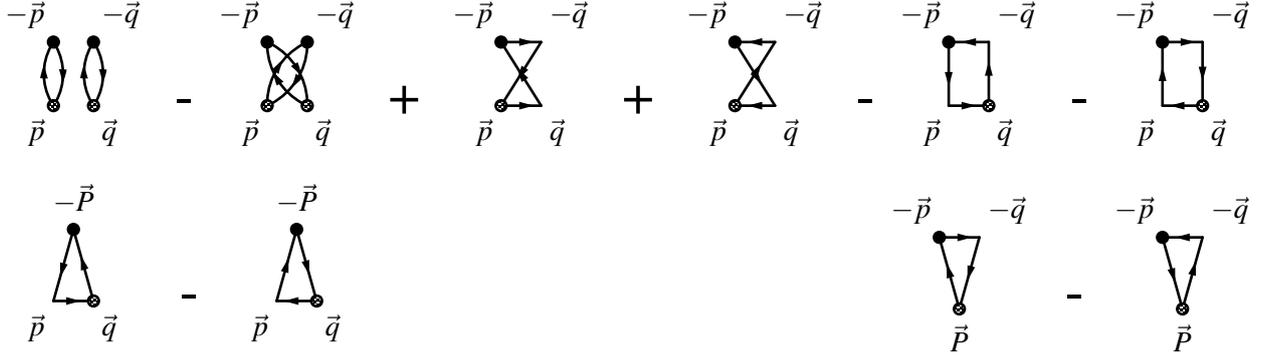
\begin{figure}[t]
\centering
\def\mydiagramwidth{15}
\def\mydiagramheight{30}
\def\mydiagramblack{\fmfv{decor.shape=circle,decor.filled=full,decor.size=2thick}}
\def\mydiagramgray{\fmfv{decor.shape=circle,decor.filled=gray50,decor.size=2thick}}
\def\mydiagramarrow{2mm}
\begin{fmffile}{fgraphs}
\vspace{0.5cm}
\begin{fmfgraph*}(\mydiagramwidth,\mydiagramheight)
\fmfset{arrow_len}{\mydiagramarrow}
\fmfbottom{i1,i2}
\fmftop{o1,o2}
\fmf{fermion,left=0.3}{i1,o1,i1}
\fmf{fermion,left=0.3}{i2,o2,i2}
\mydiagramgray{i1,i2}
\mydiagramblack{o1,o2}
\fmflabel{$\vec{p}$}{i1}
\fmflabel{$\vec{q}$}{i2}
\fmflabel{$-\vec{p}$}{o1}
\fmflabel{$-\vec{q}$}{o2}
\end{fmfgraph*}
\hfill{\huge -}\hfill
\begin{fmfgraph*}(\mydiagramwidth,\mydiagramheight)
\fmfset{arrow_len}{\mydiagramarrow}
\fmfbottom{i1,i2}
\fmftop{o1,o2}
\fmf{fermion,left=0.3}{i1,o2,i1}
\fmf{fermion,left=0.3}{i2,o1,i2}
\mydiagramgray{i1,i2}
\mydiagramblack{o1,o2}
\fmflabel{$\vec{p}$}{i1}
\fmflabel{$\vec{q}$}{i2}
\fmflabel{$-\vec{p}$}{o1}
\fmflabel{$-\vec{q}$}{o2}
\end{fmfgraph*}
\hfill{\huge+} \hfill
\begin{fmfgraph*}(\mydiagramwidth,\mydiagramheight)
\fmfset{arrow_len}{\mydiagramarrow}
\fmfbottom{i1,i2}
\fmftop{o1,o2}
\fmf{fermion}{i1,i2,o1,o2,i1}
\mydiagramgray{i1}
\mydiagramblack{o1}
\fmflabel{$\vec{p}$}{i1}
\fmflabel{$\vec{q}$}{i2}
\fmflabel{$-\vec{p}$}{o1}
\fmflabel{$-\vec{q}$}{o2}
\end{fmfgraph*}
\hfill{\huge+} \hfill
\begin{fmfgraph*}(\mydiagramwidth,\mydiagramheight)
\fmfset{arrow_len}{\mydiagramarrow}
\fmfbottom{i1,i2}
\fmftop{o1,o2}
\fmf{fermion}{i1,o2,o1,i2,i1}
\mydiagramgray{i1}
\mydiagramblack{o1}
\fmflabel{$\vec{p}$}{i1}
\fmflabel{$\vec{q}$}{i2}
\fmflabel{$-\vec{p}$}{o1}
\fmflabel{$-\vec{q}$}{o2}
\end{fmfgraph*}
\hfill{\huge-}\hfill
\begin{fmfgraph*}(\mydiagramwidth,\mydiagramheight)
\fmfset{arrow_len}{\mydiagramarrow}
\fmfbottom{i1,i2}
\fmftop{o1,o2}
\fmf{fermion}{i1,i2,o2,o1,i1}
\mydiagramgray{i2}
\mydiagramblack{o1}
\fmflabel{$\vec{p}$}{i1}
\fmflabel{$\vec{q}$}{i2}
\fmflabel{$-\vec{p}$}{o1}
\fmflabel{$-\vec{q}$}{o2}
\end{fmfgraph*}
\hfill{\huge-}\hfill
\begin{fmfgraph*}(\mydiagramwidth,\mydiagramheight)
\fmfset{arrow_len}{\mydiagramarrow}
\fmfbottom{i1,i2}
\fmftop{o1,o2}
\fmf{fermion}{i1,o1,o2,i2,i1}
\mydiagramgray{i2}
\mydiagramblack{o1}
\fmflabel{$\vec{p}$}{i1}
\fmflabel{$\vec{q}$}{i2}
\fmflabel{$-\vec{p}$}{o1}
\fmflabel{$-\vec{q}$}{o2}
\end{fmfgraph*}
\\\vspace{1.5cm}
\begin{fmfgraph*}(\mydiagramwidth,\mydiagramheight)
\fmfset{arrow_len}{\mydiagramarrow}
\fmfbottom{i1,i2}
\fmftop{o}
\fmf{fermion}{i1,i2,o,i1}
\mydiagramgray{i2}
\mydiagramblack{o}
\fmflabel{$\vec{p}$}{i1}
\fmflabel{$\vec{q}$}{i2}
\fmflabel{$-\vec{P}$}{o}
\end{fmfgraph*}
\hfill{\huge-}\hfill
\begin{fmfgraph*}(\mydiagramwidth,\mydiagramheight)
\fmfset{arrow_len}{\mydiagramarrow}
\fmfbottom{i1,i2}
\fmftop{o}
\fmf{fermion}{i1,o,i2,i1}
\mydiagramgray{i2}
\mydiagramblack{o}
\fmflabel{$\vec{p}$}{i1}
\fmflabel{$\vec{q}$}{i2}
\fmflabel{$-\vec{P}$}{o}
\end{fmfgraph*}
\hspace{8cm}
\begin{fmfgraph*}(\mydiagramwidth,\mydiagramheight)
\fmfset{arrow_len}{\mydiagramarrow}
\fmfbottom{i}
\fmftop{o1,o2}
\fmf{fermion}{i,o1,o2,i}
\mydiagramgray{i}
\mydiagramblack{o1}
\fmflabel{$-\vec{p}$}{o1}
\fmflabel{$-\vec{q}$}{o2}
\fmflabel{$\vec{P}$}{i}
\end{fmfgraph*}
\hfill{\huge-}\hfill
\begin{fmfgraph*}(\mydiagramwidth,\mydiagramheight)
\fmfset{arrow_len}{\mydiagramarrow}
\fmfbottom{i}
\fmftop{o1,o2}
\fmf{fermion}{i,o2,o1,i}
\mydiagramgray{i}
\mydiagramblack{o1}
\fmflabel{$-\vec{p}$}{o1}
\fmflabel{$-\vec{q}$}{o2}
\fmflabel{$\vec{P}$}{i}
\end{fmfgraph*}
\\\vspace{0.5cm}
\end{fmffile}
\caption{The contractions of $\pi\pi\to\pi\pi$ (top), $\pi\pi\to\rho$ (bottom-left) and $\rho\to\pi\pi$ (bottom-right), the $\rho\to\rho$ being trivial. Time flows upward from $0$ to $t$. Black dots represent an explicit summation whereas shaded dots represent a noise-noise contact. Between those dots we can have $Q$ propagators (one line) or $W$ propagator (two lines).}
\label{fig:contractions}
\end{figure}
The contraction are described in Fig.~\ref{fig:contractions}, and here we make explicit the first one :
\begin{equation}
G_{\pi\pi\rightarrow\rho}^{1st} = \sum_{j,\vec x} e^{-i\vec P\vec x}\left<Q(\vec x, t\mid\vec 0,t_s,\xi_j)W^\dag(\vec x, t\mid -\vec p,t_s\mid-\vec q,t_s,\xi_j)\gamma_5\gamma_3\right> .
\end{equation}

\section{Preliminary results}
We use the simulation setup of the Budapest-Marseille-Wuppertal collaboration \cite{BMW08, BMW-scaling} with 2 levels of HEX smearing \cite{physpoint-A, physpoint-B, physpoint-p}, featuring $N_f=2+1$ flavors of tree-level improved Clover fermions\cite{BMW-scaling} and the tree-level improved L\"uscher-Weisz gauge action\cite{gaugeaction}. We choose two simulations, for which the two lowest-lying scattering states are near the crossing.

The first point is for $m_\pi\simeq200~\mathrm{MeV}$, with $\beta=3.31$ and $a=0.116~\mathrm{fm}$ with a lattice size of $32^3\times48$. Contractions were computed in the center-of-mass frame $\vec P=(0,0,0)$, in which the $\rho$ mass nearly crosses the $\pi(0,0,2\pi/L)\pi(0,0,-2\pi/L)$ free two-pion state.

The other point has $m_\pi\simeq340~\mathrm{MeV}$, with $\beta=3.31$ and $a=0.116~\mathrm{fm}$. Here, the lattice size is $24^3\times48$. The analysis is performed in the moving frame $\vec P=(0,0,2\pi/L)$, in which the $\rho$ can nearly "decay" into $\pi(0,0,2\pi/L)\pi(0,0,0)$.

\begin{figure}[t]
\centering
\includegraphics[width=0.6\linewidth]{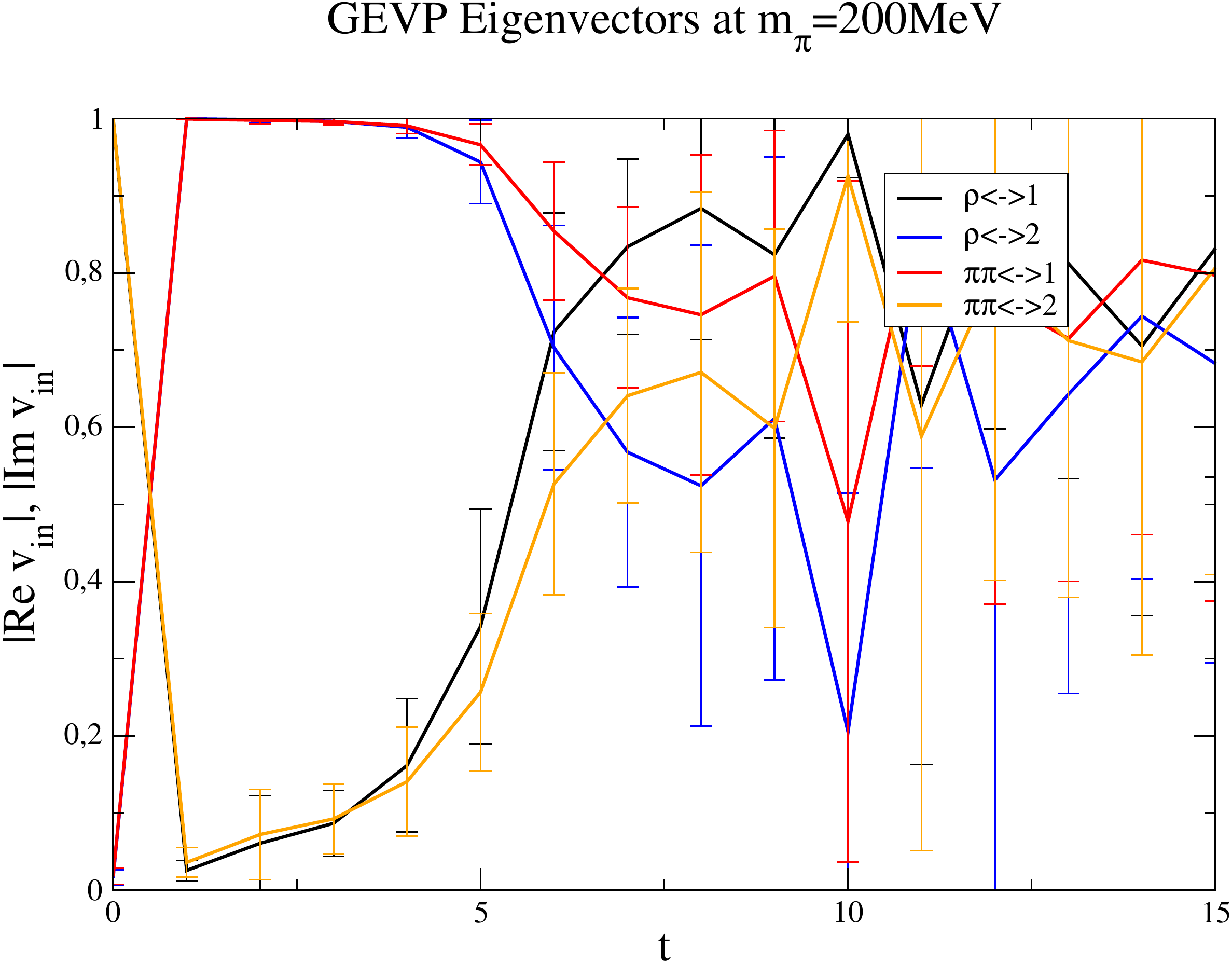}
\caption{Components of generalized eigenvectors, normalized and obtained from operators normalized so that $\langle0\mid{\cal O}(1){\cal O}(0)\mid0\rangle=1$. The eigenvectors are very sensitive to both higher-level contamination and statistical errors. For $t<6$ higher-state contamination is obvious, but for $6\le t\le 9$ or even more the eigenvectors are constant as should be.}
\label{fig:eigenvec}
\end{figure}

\begin{figure}[t]
\centering
\includegraphics[width=0.6\linewidth]{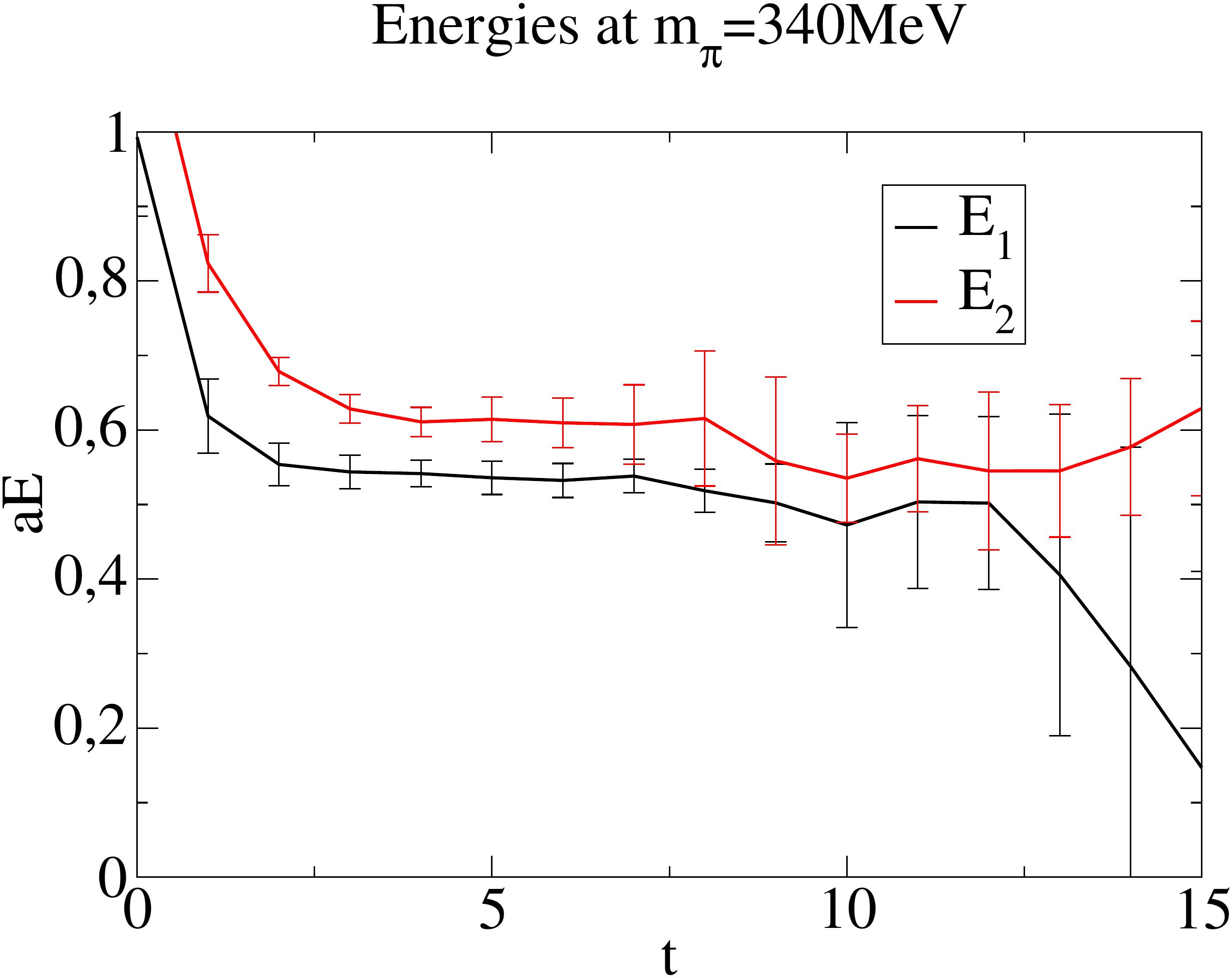}
\caption{The two energies obtained from solving generalized eigenvalue problem show a clean plateau, and their energy difference is significant. The same goes for $m_\pi=200~\mathrm{MeV}$.}
\label{fig:GEVPspectrum}
\end{figure}

Using Eq.~\ref{eq:simplecorrections} on the two energies resulting from the variational method we get (see Fig.~\ref{fig:eigenvec} and \ref{fig:GEVPspectrum}) :
\begin{eqnarray}
g = 5.5 \pm 2.9 \qquad \mathrm{for\ the}\ m_\pi\simeq200\mathrm{MeV\ point,}\\
g = 6.6 \pm 3.4 \qquad \mathrm{for\ the}\ m_\pi\simeq340\mathrm{MeV\ point,}
\end{eqnarray}
where the error is purely statistical. Combining the two results we get
\begin{equation}
g = 6.0 \pm 2.2,
\end{equation}
which is in good agreement with experimental data.

\section*{Aknowledgments}
Computations were performed using HPC resources from GENCI-[CCRT/IDRIS] (grant 52275), from FZ J\"ulich and from the DECI-5 project HADWIDTH, as well as clusters at Wuppertal and CPT. This work is supported in part by EU grants FP7/2007-2013/ERC n$^o$208740, MRTN-CT-2006-035482 (FLAVIAnet), OTKA grant AT049652 DFG grant FO 502/2, SFB-TR 55, U.S. Department of Energy Grant No. DE-FG02-05ER25681, by CNRS grants GDR n$^o$2921 and PICS n$^o$4707.

\end{document}